\documentclass[12pt]{article}
\usepackage[totalwidth=460truept,totalheight=600truept]{geometry}
\usepackage{latexsym,amssymb,amsmath,graphicx}
\usepackage[T1]{fontenc}
\usepackage{color}

\global\arraycolsep=1truept
\newcommand{\beq}{\begin{equation}}
\newcommand{\eeq}{\end{equation}}
\def\bea{\begin{eqnarray}}
\def\eea{\end{eqnarray}}

\newcommand{\ma}[1]{\mbox{$\mathcal{#1}$}}

\definecolor{darkred}{rgb}{.8,0,0}

\definecolor{darkblu}{rgb}{0,0,.8}

\begin{document}

%\hfill UB-ECM-PF-09/24

%\hfill UB-ICC-09/xx

\vskip 1.4truecm

\begin{center}
{\huge \textbf{Probing pure Lovelock gravity by Nariai and Bertotti-Robinson solutions}}

\vskip 1.5truecm

\textsl{Naresh Dadhich}

\textsl{Centre for Theoretical Physics, Jamia Millia Islamia, New Delhi 110025}

\textit{\&}

\textsl{IUCAA, Post Bag 4, Ganeshkhind, Pune 411 007, India}

{\footnotesize nkd@iucaa.ernet.in}

\vskip 1.3truecm

\textsl{Josep M. Pons}

\textit{DECM and ICC, Facultat de F\'{\i}sica, Universitat de Barcelona,\\ Diagonal 647, E-08028 Barcelona,
Catalonia, Spain.}

{\footnotesize pons@ecm.ub.edu}

\vskip 2truecm

\end{center}

\abstract{The product spacetimes of constant curvature describe in Einstein gravity, which is linear in  Riemann curvature, Nariai
metric which is a solution of $\Lambda$-vacuum when curvatures are equal, $k_1=k_2$, while it is Bertotti-Robinson metric
describing uniform electric field when curvatures are equal and opposite, $k_1=-k_2$. We probe pure Lovelock gravity by these simple 
product spacetimes and prove that the same characterization of these solutions is indeed true in general for pure Lovelock 
gravitational equation of order $N$ in $d=2N+2$ dimension. We also consider these solutions for the conventional setting of 
Einstein-Gauss-Bonnet gravity.

\section{Introduction}

It has recently been established \cite{dgj} that pure Lovelock gavity which is the $N$th order polynomial in Riemann curvature has universal
behaviour in $d=2N+1, 2N+2$ dimensions where $N=1$ is the usual Einstein gravity. For instance, it is well known that vacuum is kinematic in $3$ dimension. That is, there exists no non-trivial vacuum solution (vacuum is flat) as  $R_{ab}=0$ implies $R_{abcd}=0$ while it becomes dynamical in $4$ dimension. By universality we mean that this is true in
general for any Lovelock order $N$. That is, vacuum as defined by $R^{(N)}_{ab}=0$ is Lovelock flat, $R^{(N)}_{abcd}=0$ in critical odd
dimension $d=2N+1$ while $R^{(N)}_{abcd}$ becomes nontrivial in $d=2N+2$. Here $N$th order analogue of Riemann is as defined in \cite{d1} by the property that trace of its Bianchi derivative
vanishes and thereby yielding the divergence free $N$th order analogue of Einstein tensor, $G^{(N)}_{ab}$ which is the same as the one obtained by variation of $N$th order Lovelock action. Further there exists BTZ black hole analogue \cite{btz} in all odd $d=2N+1$ dimension. It should be noted that this happens only for pure Lovelock and not for Einstein-Lovelock gravity.

It also turns out that static
pure Lovelock black hole solution asymptotically goes over to the corresponding Einstein solution \cite {d2,dpp} even though the equation is
free of Einstein term. Not only that its temperature and entropy bear the universal relationship to event horzon radius \cite{dpp1}.
That is, entropy always goes as $r_h^2$ or $A^{1/N}$ where $A$ is area of black hole in $d=2N+2$ dimension.

It has therefore been proposed in Ref. \cite{prague} that the vacuum gravitational equation in dimension $d = 2N+1, 2N+2$ is 
\beq 
G^{(N)}_{ab}=-\Lambda g_{ab}. 
\eeq
Note that in the conventional Einstein-Lovelock equation, there a is sum over all the terms $\leq N$. There higher order terms are thought to be higher order corrections to Einstein gravity and this viewpoint is motivated by string theory in which one loop corrections do yield Gauss-Bonnet term along with much else. Here in contrast we would like to consider classical gravity in higher dimensions on its own footing and not as corrections to Einstein gravity. Then the question is: Is it possible to carry over the property that gravity is kinematic in all odd dimensions? That means this is a universal gravitational property. To that there is unique answer, pure Lovelock gravity. This is our main motivation for the above equation. Of course it should be a second order equation, that it is, so that initial value problem is well defined to give unique evolution. Another desirable feature is to have is that asymptotically it should approximate to Einstein solution. That is exactly what happens for static vacuum solutions of the above equation \cite{d2,dpp}. Not only that it has been recently also shown that bound orbits around a static black hole can exist in all even dimensions $d=2N+2$ only for pure Lovelock gravity and in no other theory \cite{dgj1}. That is bound orbits exist for Einstein gravity only in $4$ dimension and in none else while pure Lovelock gravity they do in all even dimensions. It is as remarkable a result as Bertrand's theorem of classical mechanics. 

It would thus be pertinent to probe this equation for various situations. As has already been pointed out that static
black hole solutions of this equation asymptotically tend to the Einstein Schwarzschild-dS/AdS in $d = 2N+1, 2N+2$ dimension \cite{d2,dpp} even
though there is no Einstein term in the equation. In this paper we wish to probe this equation by employing the ansatz of product spaces of
constant curvature. We would consider spacetimes ${\ma M}^d=M^2\times \Sigma^{d-2}$, product of two spaces of constant curvature (the former is
dS/AdS/flat while the latter is a sphere or hyperbolic manifold manifold of constant radius, or Euclidean space). For this ansatz in $4$ dimension, there are two
well known solutions described by Nariai metric \cite{nar} representing $\Lambda$-vacuum when curvatures are equal, $k_1=k_2$ and
Bertotti-Robinson \cite{ber,rob} metric describing uniform electric field when curvatures are equal and opposite, $k_1=-k_2$. It turns out that
Nariai metric, unlike dS/AdS is not conformally flat, while Bertotti-Robinson (BR), unlike other Einstein-Maxwell solutions, is indeed conformally
flat. We would now like to study their analogues in pure Lovelock gravity to establish their universal characterization. That is, it is
$k_1=k_2$ for Nariai and $k_1=-k_2$ for BR in all $d=2N+2$ dimension. In the process we have also discovered a property of Lovelock gravity that
pure Lovelock vacuum solutions exist for $N>1$ even when one of the curvatures $k_1, k_2$ vanishes. This happens because $R^{(N)}_{ab}$ is
homogeneous of degree $N$ in $R_{ab}$ involving number of terms and they could cancel out each-other to make it zero. This is why even when some
of Ricci components are constants, $R^{(N)}_{ab}$ could vanish. At any rate these are curious and interesting pure Lovelock vacuum solutions.

The paper is organised as follows. In the next section we set up the Lovelock framework for the general case that includes both Nariai and BR
solutions which is followed by establishment of their universal characterization in $d=2N+2$ dimensions as well as the solutions in Einstein-Gauss-Bonnet setting. We conclude with a discussion. 

%Next we shall discuss some new solutions, for instance it is possible to have a non-trivial solution even in the critical odd dimension
%$d=2N+1$ with a finetuning between $\Lambda$ and electric field. It is followed by Nariai and BR solutions in Einstein-Gauss-Bonnet gravity and
%discussion of interesting pure Lovelock vacuum solutions sourced by one of the curvatures which remains undetermined. We conclude with a
%discussion.

\section{Nariai-Bertotti-Robinson solutions for the general Lovelock Lagrangian}

The general Lovelock Lagrangian in $d$ dimensions is a polynomial of the Riemman tensor \beq \label{LovLag} \frac{1}{2\kappa_d^2}\sum_{p=0}^N
\frac{(d-2p-1)!}{(d-1)!} c_p {\ma L}_p, \eeq with $ N \leq \left\lfloor\frac{d-1}{2}\right\rfloor$ to have nontrivial dynamics
(For $ d=2 N $ the term ${\ma L}_N$ is an Euler  density, which is a total divergence), and where ${\ma L}_p$ is a monomial of order $p$,
\beq\label{lovelocklag} {\ma L}_p=\frac{1}{2^p}\sqrt{-g}\,\delta^{\mu_1\cdots \mu_p\nu_1\cdots \nu_p}_{\rho_1\cdots \rho_p\sigma_1\cdots
\sigma_p}R_{\mu_1\nu_1}^{\phantom{\mu_1}\phantom{\nu_1}\rho_1\sigma_1}\cdots R_{\mu_p\nu_p}^{\phantom{\mu_p}\phantom{\nu_p}\rho_p\sigma_p}, \eeq
where $\kappa_d := \sqrt{8\pi G_d}$. We take units such that $2 \kappa_d^2=1$. The $\delta$ symbol is defined by \bea \delta^{\mu_1\cdots
\mu_p\nu_1\cdots \nu_p}_{\rho_1\cdots \rho_p\sigma_1\cdots \sigma_p}&:=&(2p)!\,\delta^{\mu_1}_{[\rho_1}\cdots \delta^{\nu_p}_{\sigma_p]}, \eea
so that it takes values $0$ and $\pm 1$. Note that $c_p$ is a coupling constant with dimension $(l^2)^{p-1}$. The factorial coefficients in (\ref{LovLag}) are aimed
to match for the coefficients $c_p$ of the standard results in the literature. In our conventions $c_0=-2\Lambda$,  where $\Lambda$ is the cosmological constant, and to describe the
Einstein-Hilbert Lagrangian we must use $c_1=(d-1)(d-2)$.

\vspace{4mm}

We will look for solutions under the general ansatz
\beq
ds^2 = -A(r)\,dt^2 + B(r)\,dr^2 + C(r)\,d\Omega^2_{(d-2)}\,,
\label{construnc}
\eeq
where $\Omega_{(d-2)}$ is the $(d-2)$-sphere. This ansatz is a consistent truncation; i.e. such that the truncated equation of motion (EOM) of the original  Lagrangian is the same as the EOM of the truncated Lagrangian of the theory, because there is no source in the EOM of the original Lagrangian to enforce the appearance of non-diagonal terms for the metric nor to modify the spherical symmetry of the $d-2$ dimensional sector. Under this anzatz, the only nonvanishing components of the Riemman tensor are:
\bea
{R_{01}}^{01}&=& \frac{A(r) A'(r) B'(r)+B(r) \Big(A'(r)^2-2 A(r)
   A''(r)\Big)}{4 A(r)^2 B(r)^2} =:D(0,1)\,,\nonumber\\
{R_{0a}}^{0a}&=&-\frac{A'(r) C'(r)}{4 A(r) B(r) C(r)}=:D(0,a)\,,\nonumber\\
{R_{1a}}^{1a}&=&\frac{C(r) B'(r) C'(r)-2 B(r) C(r) C''(r)+B(r)
   C'(r)^2}{4 B(r)^2 C(r)^2}=:D(1,a)\,,\nonumber\\
{R_{ab}}^{ab}&=&\frac{1}{C(r)}-\frac{C'(r)^2}{4 B(r) C(r)^2}=:D(a,b)\,,\quad  (a\neq b)\,,
\label{riem}
\eea
with $a,b = 2,3,...,d-2$. %indices ($a\neq b$) running the angular coordinates of the sphere: $a,b= 2,3,\dots,d-1$. Notice that the Riemman components do not depend on the specific $a,b$ indices nor of the spacetime dimension $d$.
 The density factor becomes (up to the volume of the $(d-2)$-sphere, here irrelevant)
$$
\sqrt{-g}\to\sqrt{A(r) B(r)} C(r)^{\frac{d-2}{2}}\,.
$$
Plugging these components into (\ref{lovelocklag}) we obtain a Lagrangian in terms of the fields $A(r),
B(r),C(r)$. Defining
$${\ma L}_{{}_T}(d,p)=\frac{(d-2p-1)!}{(d-1)!} {\ma L}_p|_{\rm trunc.}$$ we obtain
 \bea
{\ma L}_{{}_T}(d,p)&=& \frac{\sqrt{-g}}{(d-2p)(d-1)}\Big(D(a,b)^{p-2} \Big(2p\, D(0,1) D(a,b) \nonumber\\
&+&4p(p-1)  D(0,a)
   D(1,a)
+ 2p (d-2 p) D(a,b)
   (D(0,a)+D(1,a))\nonumber\\
&+&(d-2 p-1)
   (d-2 p)D(a,b)^2 \Big) \Big)\,.
\label{trunclag}
\eea
Now the truncation of (\ref{LovLag}) becomes
$${\ma L}_{{}_T}=\sum_{p=0}^N c_p {\ma L}_{{}_T}(d,p)\,.$$

A source of constant electric field in the radial direction can be easily acommodated
 in the framework (\ref{construnc}). The Maxwell Lagrangian
$-\frac{1}{4}F_{\mu\nu}g^{\mu\rho}g^{\nu\sigma}F_{\rho\sigma}$ becomes, for $F_{01} =E$,
\beq
{\cal L}_{\!{}_M} = \sqrt{-g}\frac{E^2}{2\, A(r)B(r)}\,.
\label{lm}
\eeq
where $E^2$ is just the square of $E$.

\vspace{4mm}

Had we considered an hyperbolic manifold instead of a sphere in (\ref{construnc}), the only  difference would have been that $C(r)\to -C(r)$ in
(\ref{riem}).

Let us now specialize the ansatz of product spaces and consider ${\ma M}^d=M^2\times \Sigma^{d-2}$ where $\Sigma^{d-2}$ is of
constant curvature. Then the EOM demand that $M^2$ is also of constant curvature, $dS/AdS/{\rm flat}$ \cite{d3}. We then write the metric, \beq
ds^2 = -(1-k_1 r^2)\,dt^2 + \frac{1}{1-k_1 r^2}\,dr^2 \pm \frac{1}{k_2}\,d\Sigma^2_{(d-2)}, \label{construnc2} \eeq where in the last term if
$\Sigma_{(d-2)}$ is the sphere $S_{(d-2)}$ we will use the plus sign and $k_2>0$, whereas if $\Sigma_{(d-2)}$ is the hyperbolic manifold $H_{(d-2)}$ we
will use the minus sign and $k_2<0$. In both cases the EOM take the same form \beq \sum_{p=0}^N (d-2p-1) c_p\, k_2^p - \frac{(d-1)}{2} E^2 =0\,,
\label{polk2} \eeq and \beq
 (2\sum_{p=1}^N p\,
c_p\, k_2^{p-1})\,k_1 +\sum_{p=0}^N (d-2p-1)(d-2p-2) c_p\, k_2^{p}+\frac{(d-1)(d-2)}{2} E^2=0\,.
\label{sec-eq-real}
\eeq
For generic values of the coefficents one can isolate $k_1$,
\beq
 k_1 = -\frac{\sum_{p=0}^N (d-2p-1)(d-2p-2) c_p\, k_2^{p}+\frac{(d-1)(d-2)}{2} E^2}{ 2\sum_{p=1}^N p\,
c_p\, k_2^{p-1}}\,, \label{sec-eq}\eeq with $k_2>0$ for the spherical case and $k_2<0$ for the hyperbolic manifold case. Actually, the case $k_2=0$ is
also described by these equations (\ref{polk2}) and (\ref{sec-eq-real}). In fact, $k_2=0$ means that the sphere or the hyperbolic manifold has become
Euclidean space (the curvature radius becoming infinite), and the ansatz (\ref{construnc2}) is no longer convenient. It must be modified with
$dE_{(d-2)}^2$ for the last term. Then one can check that the EOM for the Euclidean case coincide with the EOM (\ref{polk2}) and
(\ref{sec-eq-real}) when $k_2=0$, that is, $E^2+4\Lambda=0$, and  $(d-1)(d-2)(E^2-4\Lambda)+ 4\, c_1 k_1 =0$.

\vspace{4mm}

Summarizing, the EOM (\ref{polk2}) and (\ref{sec-eq-real}) are  general for the ansatz ${\ma M}^d=M^2\times \Sigma^{d-2}$, with $\displaystyle M^2= -(1-k_1
r^2)\,dt^2 + \frac{1}{1-k_1 r^2}\,dr^2$, and $\Sigma^{d-2}$ being the constant curvature surface: sphere, hyperbolic manifold or Euclidean space.

\vspace{4mm}

The $\Lambda$ vacuum Nariai solution for the Einstein-Lovelock gravity has been  considered in \cite{maeda} and it follows from (\ref{polk2}) and
(\ref{sec-eq}) in the limit $E=0$. By eliminating $E$ or $\Lambda$, we can also write % expression for $k_1$ can be obtained by using eq.(\ref{polk2})  eliminate $E^2$ in eq.(\ref{sec-eq}) and we get,
\beq
k_1=-\frac{\sum_{p=0}^N (d-2p-1)(d-p-2) c_p\, k_2^{p}}{\sum_{p=1}^N p\, c_p\, k_2^{p-1}}\,.
\label{solk1}
\eeq
or
%One could also use eq.(\ref{polk2}) to eliminate $c_0$ in eq.(\ref{sec-eq}), then
%we obtain
\beq k_1= k_2\Big( (d-2N-1) + 2\,\frac{\sum_{p=1}^{N-1} p(N-p)\, c_p\, k_2^{p-1}}{\sum_{p=1}^N p\, c_p\, k_2^{p-1}} \Big) -\frac{(d-1)(d-2) E^2
}{2\sum_{p=1}^N p\, c_p\, k_2^{p-1}} \label{solk1-2} \eeq The remaining EOM is that of the electromagnetic field. It is straighforward to verify
that EOM for the Maxwell field $\partial_\mu(\sqrt{-g}F^{\mu\nu})=0$ is trivally satisfied for  constant $F_{01} =E$. This is the general
electrovac Einstein-Lovelock solution (with sum over $c_p$) which is the electromagnetic (Bertotti-Robinson) generalization of Nariai
solution obtained in \cite{maeda}.

In the next section, we shall specialize to pure Lovelock gravity and then establish universality of these solutions in $d=2N+2$ dimension.

\section{Universal characterization of Nariai and Bertotti-Robinson solutions and other cases}

For the Lovelock solutions we have $c_0=-2\Lambda$ and only $c_N \neq 0$. To be specific, we will consider $c_N>0$,
%  without any loss of generality we set $c_N=1$,
then  eq.(\ref{polk2}) gives \beq (d-2 N-1)\, c_N\, k_2^N  = 2 (d-1)(\Lambda+\frac{E^2}{4}) \label{pure1} \eeq and eq.(\ref{sec-eq}), \beq
k_1=-\frac{(d-2 N-1)(d-2 N-2)}{2N}k_2 -\frac{(d-1)(d-2)}{2 N\, c_N\, k_2^{N-1}}(\frac{E^2}{2}-2\Lambda)\,. \label{pure-sec-eq} \eeq
Alternatively, using eq.(\ref{pure1}) to eliminate either $E^2$ or $\Lambda$, $k_1$ can also be written as \bea
k_1&=&-\frac{ -2(d-1)(d-2)\Lambda +(d-2N-1)(d-N-2) \, c_N\, k_2^N}{ N\, c_N\,  k_2^{N-1}}\nonumber\\
&=& (d-2N-1) k_2  -\frac{(d-1)(d-2) E^2}{2 N\, \, c_N\, k_2^{N-1}}\,.
\label{pure2}
\eea

 %\subsection{2N+2}
Let us now specialize to $d=2N+2$ and combine $k_1$ and $k_2$ together in one equation. After some simple manipulations we write
\bea\label{even1}
k_1+k_2 &=& 4\Lambda \, (\frac{2N+1}{ c_N})\, k_2^{1-N} \\
k_1-k_2 &=& -E^2 \,(\frac{2N+1}{c_N})\, k_2^{1-N}\label{even2}
\eea
and therefore
\beq
(k_1+k_2)E^2 = -4(k_1-k_2)\Lambda \,.
\label{single}
\eeq

The above equation clearly demonstrates the universal characterization of Nariai with $k_1=k_2$ and Bertotti-Robinson with $k_1=-k_2$
solutions in all $d=2N+2$ dimension. Eq.(\ref{single}) remains true for all $d=2N+2$ dimension. We get the same relations between the two
curvatures for the generalized solutions in $d=2N+2$ dimension as that for the original $4$-dimensional Nariai and Bertotti-Robinson solutions.
For the Nariai,  it is $k_1=k_2$ and $E=0$ and $\Lambda$ is given by
\beq
\Lambda = \frac{c_N}{2(2N+1)} k_1^N
\eeq
while for Bertotti-Robinson, it is $k_1=-k_2$ and $\Lambda=0$ and $E$ is given by
\beq
E^2 = \frac{2c_N}{2N+1} k_1^N.
\eeq
This means the Lovelock generalization of these solutions retains the characterizing relation intact while $\Lambda$ and $E^2$ are proportional
to $N$th power of the curvature.

\vspace{4mm}

The critical condition is $E^2\ge 0$ which requires $k_2^{N-1}(k_2-k_1)\geq 0$ and we have the
following cases:

(a) $k_2>0$. Then $k_1\le k_2$ and $\Lambda\ge 0$ for $k_2\ge|k_1|$ and vice versa for $k_2\le|k_1|$. \\

(b) $k_2<0$. Then if $N$ is odd, $k_1<0$, $|k_1|\ge |k_2|$, and $\Lambda< 0$, and if $N$ is even, then for $k_1 \ge |k_2|$, $\Lambda\le 0$ and
for $|k_1| \le |k_2|$, $\Lambda\ge 0$.
 \\

(c) $k_2=0$. This is a trivial case because curvature is non-trivial in $2$-space for which Einstein tensor is identically zero. 

%  \\

(d) $k_1=0$. Then $\displaystyle E^2=4\Lambda= (\frac{ c_N}{2N+1})\,k_2^N$ and so $k_2>0$ or $N$ even. The case with $k_2>0$ also generalizes
another Plebanski-Hacyan solution in $d=4$, \cite{pleb}, \cite{pod}. The case, $k_1=0$ and $k_2>0$, has also been studied as a string
dust distribution where $\rho=-p_r=k_2$ as an electrogravity dual of flat spacetime \cite{d3} without $\Lambda$ and $E$.

%\subsection{d=2N+1}
For $d=2N+1$, we have
\beq
 E^2=-4\Lambda = - \frac{c_N k_1 k_2^{N-1}}{2N-1},  \, \, \, \Lambda<0 \,.
 \label{2nplus1}
\eeq

This equation signifies an interesting situation that spacetime is non-trivial only when both $E$ and $\Lambda$ either non-zero or zero,
one alone cannot be zero. For $E^2>0$, if $k_1<0$, $k_2>0$ or $N$ odd, while for $k_1>0$, $N$ even and $k_2<0$. If we further assume $k_1+k_2=0$
which is the condition for vanishing of the Weyl curvature (conformal flatness), $E^2 = -4\Lambda = c_Nk_1^N/(2N-1)$. 

There also occurs an interesting vacuum solution for $N>1$ when $k_1=0$ and  $k_2\neq0$ for which both $E$ and $\Lambda$ vanish. This is a pure Lovelock vacuum spacetime which occurs only in odd $2N+1$ dimension. What we have here is a product of Minkowski $2$-spacetime with odd dimensional constant curvature space. It  is thus constant curvature $k_2$ which is responsible for this pure Lovelock vacuum.   \\

Now we analyze with more detail some Nariai-Bertotti-Robinson solutions and finally focus on the Einstein-Gauss-Bonet case.

\subsection{ $N=1$}

In this case, analysis of eq.(\ref{pure1}) and eq.(\ref{pure-sec-eq}) gives
%\bea
%{\rm sign}\,( k_2 )&=& {\rm sign}\,(\Lambda+\frac{E^2}{4})\nonumber\\
%{\rm sign}\,( k_1 )&=& {\rm sign}\,(\Lambda-(d-3)\frac{E^2}{4})\,,
%\eea
%and we infer
\bea
\Lambda\ge (d-3)\frac{E^2}{4}\ &\Longleftrightarrow&\ k_2>0,\ k_1\ge 0\nonumber\\
-\frac{E^2}{4}<\Lambda<(d-3)\frac{E^2}{4}\ &\Longleftrightarrow&\  k_2>0,\ k_1<0\nonumber\\
\Lambda\le -\frac{E^2}{4}\ &\Longleftrightarrow&\  k_2\le 0,\ k_1<0\,. \eea Note also that the Minkowski background ($k_1=k_2=0$) is only attainable
for $\Lambda=E^2=0$.)

%Case (i) $N=1$: If $k_2>0$, then $k_1\ge\le 0$, $k_2\ge\le \mod{k_1}$ and $\Lambda %\ge\le 0$ while if $k_2<0$, then $k_1<0$, $\mod{k_1}>\mod{k_2}$ and $\Lambda<0$. If %$k_2=0$, $k_1<0$ and $\Lambda<0$, and if $k_1=0$ then $k_2>0$ and $\Lambda>0$.

\subsection{ $N>1$ in general}
Let us consider first the case $\Lambda+\frac{E^2}{4}=0$ and $d\neq2N+1$. Eq.(\ref{pure1}) imposes then $k_2=0$, and consistency with
eq.(\ref{pure-sec-eq}) demands $\Lambda=E=0$ with $k_1$ remaining arbitrary. This would be pure Lovelock vacuum solution as mentioned
earlier. When $\Lambda+\frac{E^2}{4}\neq 0$, we have the following general relation between the curvatures,
\beq
 k_1=-\frac{d-2N-1}{N}\Big(\frac{(d-N-2)E^2-4N\Lambda}{E^2+4 \Lambda}\Big)k_2\,.
\label{pure4}
\eeq
Note that the relations (\ref{even1}) (\ref{even2}), follow from it for $d=2N+2$.

%Keeping with $\Lambda+\frac{E^2}{4}\neq 0$, eq.(\ref{pure1}) allows for a straightforward analysis of different cases for $k_1$, $k_2$. Instead,
%we  will discuss the generalization of the Nariai and Bertotti-Robinson solutions.

\subsection{Nariai and Bertotti-Robinson solutions for $d\neq 2N+2$}

Now we specialize to the case where either $E$ vanishes (Nairai solutions) or $\Lambda$ vanishes  (Bertotti-Robinson solutions). One can rewrite eq.(\ref{pure4}) as
\beq
(k_1+ \frac{(d-2 N-1)(d-N-2)}{N}\, k_2)\,E^2= -4  \,(k_1- (d-2 N-1)\, k_2)\,\Lambda\,,
\eeq
which is a generalization of eq.(\ref{single}). We obtain:
\bea
\underline{\rm Nariai\ solution}&:& \quad  k_1= (d-2 N-1)\, k_2\,,\nonumber\\
\underline{\rm Bertotti\!-\!Robinson\ solution}&:& \quad k_1=- \frac{(d-2 N-1)(d-N-2)}{N}\, k_2\,. \eea

As before $\Lambda$ and $E^2$ would go as $k_1^N$.

\subsection{Conformally flat solutions for pure Lovelock}
\label{nbr-2}

It turns out that for the metric (\ref{construnc2}), the Weyl curvature is proportional to $k_1+k_2$ and hence it can only vanish when
$k_1+k_2=0$ which was the condition for the Bertotti-Robinson soultion in general for $d=2N+2$ describing a uniform electric field.

%in the $4$-dimensional spacetime.

 Using eq.(\ref{pure4}) we obtain
%\beq {\rm \underline{Conformally\ flat\spacetime}}: \
\beq (k_1+k_2=0)\  \Leftrightarrow\ \Lambda= \frac{(d-2N-2)(d-N-1)}{N(d-2N)}\frac{E^2}{4}\,. \label{finet} \eeq

Clearly $\Lambda=0$ for $d=2N+2$ characterizing the Bertotti-Robinson solution. On the other hand for $d=2N+1$ as noted earlier,
$E^2+4\Lambda=0$ irrespective of $k_1+k_2=0$ or not.
% Unless both $E$ and $\Lambda$ vanish, it would always be conformally flat.

% It is clear from
%eq.(\ref{pure4}) that
 %when $E^2+4\Lambda=0$, then $d=2N+1$ and it would be conformally flat if in addition $k_1+k_2=0$, then in
%view of eq.(\ref{2nplus1}), either $k_1<0$ or $N$ is even so as to keep $E^2>0$.

\subsection{Nariai and Bertotti-Robinson solutions for Ein\-stein-Gauss-Bonnet gravity}

Let us apply the general equations (\ref{polk2}) and (\ref{sec-eq}) to the case $N=2$ in arbitrary $d\geq 5$ dimensions. Taking $c_0=-2
\Lambda,\ c_1= (d-1)(d-2)$, and $c_2 = (d-1)(d-2)(d-3)(d-4)\, \alpha$, the Lagrangian (\ref{LovLag}) with the Maxwell term is
\beq {\ma L}=
\sqrt{-g} \Big(-2 \Lambda + {R} + \alpha\,({R}^2 -4 {R_\mu}^\nu{R_\nu}^\mu + {R_{\mu\nu}}^{\rho\sigma}{R_{\rho\sigma}}^{\mu\nu})
-\frac{1}{4}F_{\mu\nu}g^{\mu\rho}g^{\nu\sigma}F_{\rho\sigma}\Big)\,.
\eeq

Then the EOM give,
$$
k_2=\frac{E^2+4\Lambda}{(d-2)(d-3)\pm\sqrt{(d-3) (d-2)
   \left((d-2)(d-3)+2 (d-5) (d-4)(E^2+4\Lambda)
   \alpha \right)}}
$$
which requires
\beq
(d-2)(d-3)+2 (d-5) (d-4)(E^2+4\Lambda)
   \alpha\geq 0
\label{realk} \eeq for $k_2$ to be real, and  $k_1$ is given by
$$
k_1=
   -\frac{2 (d-4) (d-3)
   k_2 ((d-6) (d-5)
   k_2\, \alpha
   +1)+E^2-4 \Lambda }{8
   (d-4) (d-3) k_2\,
   \alpha +4}
$$

Let us take $d=5$, then the above equations can be solved for $E$ and $\Lambda$,
%yield
%\beq E^2+4\Lambda = 12k_2, \eeq
%\beq E^2-4\Lambda = - 4(k_1+k_2 + 4\alpha k_1k_2).
%\eeq
%We can then solve for $E$ and $\Lambda$,
$$
E^2 = 2(2k_2-k_1-4\alpha K_1k_2),\qquad
2\Lambda = k_1 + 4k_2 + 4\alpha k_1k_2.
$$
Now if $E=0$, then
$$
k_1(1 + 4\alpha k_2) = 2k_2,\quad  {\rm and}\quad \Lambda = 3 k_2.
$$
On the other hand, if $\Lambda=0$, then
$$
k_1(1 + 4\alpha k_2) = -4k_2,\quad   {\rm and}\quad E^2 = 12 k_2.
$$

\vspace{4mm}

Now since we have the additional GB coupling parameter, $\alpha$, which allows much room for manipulation to have various possibilities of
conformally flat solutions by appropriately fine tuning  $\Lambda$ and $\alpha$. The conformally flat condition, $k_1+k_2=0$, implies a
relationship among the parameters $\Lambda,\,\alpha$ and $E^2$.
Let us consider $d=6$ and then we write
% and we write
%$$
%\alpha  =\frac{2(d-2)^2(d(d-7)+14)((d-4)E^2-4\Lambda) }
%{(d-3)(d-4)\Big((d-3)(d-6)E^2-8(d-4)\Lambda   \Big)^2}
%$$
%which satisfies (\ref{realk}) and
%$$
%k_2=-k_1= \frac{8(d-4)\Lambda-(d-3)(d-6)E^2}{2(d-2)(d(d-7)+14)}.
%$$
%Note that the requirement $k_2\neq 0$ guarantees that $\alpha$ is well defined. In %particular we see that even in absence of the electric
%field, a conformally flat solution exists with
%$$
%\alpha  =-\frac{(d-2)^2(d(d-7)+14) }
%{8(d-3)(d-4)^3\Lambda}\,.
%$$
%For $d=6$ we obtain,
%for the conformally flat condition
$$2 \Lambda + 3 \Lambda^2\alpha - E^2=0\,,
$$
and
$$
k_2=-k_1=\frac{\Lambda}{4}\,.
$$
Notice the critical case $\Lambda\alpha=-\frac{2}{3}$, which yields a conformally flat solution and yet $E=0$.  This is in contrast to the
Bertotti-Robinson solution which described uniform electric field in Einstein gravity.

\section{Discussion}

We have studied here generalizations of Nariai and Bertotti-Robinson solutions in pure Lovelock gravity and the main result is that their
characterization is universal; i.e. the same as for Einstein gravity in $4$ dimension. That is, for $d=2N+2$, Nariai is always characterized by
the condition, $k_1=k_2$ while it is $k_1=-k_2$ for Bertotti-Robinson. This is the same characterization as in $4$ dimension. It is
important that this happens only in even $d=2N+2$ dimensions and not in any other. This is what makes this result non-trivial despite its
very special simple metric ansatz. This is in line with the general view that gravitational dynamics is essentially universal in all odd/even dimensions $d=2N+1, \, 2N+2$ \cite{dgj}. In $d=2N+1$ dimension, the solution exists only for $E^2=-4\Lambda$. For particular
cases of one of curvatures vanishing, solutions are generalizations of their $4$-dimensional Plebanski-Hacyan analogues \cite{pleb}. There also exists generalized Nariai-Bertotti-Robinson solution with both $E$ and $\Lambda$ present.\\

\section*{Acknowledgments}
JMP acknowledges partial financial support from projects FP2010-20807-C02-01,
2009SGR502 and CPAN Consolider CSD 2007-00042.

\bibliographystyle{elsarticle-num}

\end{document}